\documentclass[epj]{svjour}
\usepackage{graphics}
\begin{document}
\title{Transient dynamics of linear quantum amplifiers}
\author{Sabrina Maniscalco\inst{1} \and Jyrki Piilo\inst{1} \and Nikolay Vitanov \inst{2} \and Stig Stenholm \inst{3}}
%
%
\institute{School of Pure and Applied Physics, University of
KwaZulu-Natal, Durban 4041, South Africa \and Department of
Physics, Sofia University, James Boucher 5 Boulevard, 1164 Sofia,
Bulgaria \and Laser Physics and Quantum Optics, Royal Institute of
Technology (KTH), Alba Nova, Roslagstullsbacken 21, SE-10691
Stockholm, Sweden}
\date{Received: date / Revised version: date}
%
\abstract{ The transient dynamics of a quantum linear amplifier
during the transition from damping to amplification regime is
studied. The master equation for the quantized mode of the field
is solved, and the solution is used to describe the statistics of
the output field. The conditions under which a nonclassical input
field may retain  nonclassical features at the output of the
amplifier are analyzed and compared to the results of earlier
theories. As an application we give a dynamical description of the
departure of the system from thermal equilibrium.
} 
\maketitle
\section{Introduction}
\label{intro}
The master equation describing linear amplification
or gain has been a prototype for discussing many questions in
Quantum Optics. It was derived by the elimination of an unobserved
environment using what has been termed a Born-Markov approximation
\cite{mandel}. Thus its properties were mainly determined by
physical considerations, but it arrived at a form later to be
shown to be the consistent generator of dissipative time evolution
in quantum theory; thus it is of the Lindblad form
\cite{Lindblad}.

The master equation for linear amplification was earlier
represented as the generic model for an optical amplifier or
attenuator \cite{glauber}. It also describes the onset of laser
oscillations until the time when nonlinear saturation starts to
affect the behavior. In the trapped ion context, when the ion trap
potential is regarded in a harmonic approximation, the cooling by
lasers may be considered as an attenuation described by the same
equation \cite{blatt}.

Its advantage is that it is exactly solvable, which allows us to
follow the onset of gain or the damped approach to a steady state.
The exact solution also allows us to evaluate the noise properties
exactly and investigate the fading of nonclassical features of the
initial state.

In all applications so far, the amplifying and attenuating
coefficients of the equation have been regarded as constants. This
corresponds to the assumption that the population inversion is
instantaneously reached and the evolution starts from an initial
state experiencing no previous evolution. Such theory, however,
does not describe the transient dynamics of the linear amplifier,
i.e. the dynamics when the pumping field is switched on or off. In
this paper we generalize the previous theory of linear amplifiers
to the case in which a smooth onset of amplification or
attenuation takes place, and hence the amplifying and attenuation
coefficients are time dependent. We solve the master equation with
time dependent coefficients in terms of the characteristic
function \cite{PRAsolanalitica,misbelief} and we use the solution
to describe the transient dynamics of the linear quantum
amplifier, one of the most widely used and common devices in
quantum optics.

We choose to consider a situation where the gain medium is
switched smoothly from an attenuating regime to an amplifying one.
This takes place within a time interval centered at some definite
time, before which we have a damped situation. Thus the initial
gain, which is normalized to unity of course, decreases first
until the amplification coefficient changes sign and gain starts
to grow. The model allows us to follow the solution through this
point, and we can see how the time dependence affects the noise
properties and the possibility to retain initially imposed
nonclassical features of the system state. As we may expect, the
situation is more complicated than the simple constant coefficient
case. It is, however, possible to retain earlier results on
amplifier added noise and quantum cloning limits by considering
the appropriate limiting cases.

Theoretical works on optical transients of physical phenomena
which cause amplification of light have received in the past a
huge deal of attention \cite{opttransients}. To the best of the
authors' knowledge, however, this is the first analytic
description of the transient of phase-insensitive quantum linear
amplifiers. Therefore the results presented in this paper give a
clear contribution to the fundamental research in the theory of
lasers and optical amplifiers since every linear amplifier or
laser undergoes a transient behavior before stabilizing.

Our results can be directly applied to describe transients in
technological applications based on linear amplifiers. Optical
linear amplifiers are essential components of state-of-the-art
optical networks. In order to attain best performances of the
networks, however, it is crucial to analyze the behavior of linear
amplifiers during power transients causing fast switching on and
off of the amplifiers \cite{applicationLA}. Although the linear
amplifiers currently used in optical networks do not need to
operate at the quantum level, the recent development of quantum
technologies such as quantum communication, quantum cryptography
and quantum computation, is based on the implementation of
networks containing nanodevices and nanocomponents operating at
the quantum level.

Recently a scheme for optimal cloning of coherent states with
phase-insensitive linear amplifiers and beams splitters has been
proposed \cite{Braunstein01}. Recent advances in the field of
nanoelectromechanical systems have paved the way to the
realization of experiments close to achieve the quantum limited
detection and amplification. In \cite{LaHaye04}, position
resolution very close to the quantum limit is obtained,
demonstrating the near-ideal performance of a single-electron
transistor as a linear amplifier. Another recent application of
linear amplifiers consists in a method for reconstructing a
multimode entangled state \cite{Zubairy03}. Also quasiprobability
functions have been shown to be measurable via direct
photodetection of a linear amplified field \cite{Kim97}.

The paper is structured as follows. Section II summarizes the
results of earlier investigations for easy comparison with the
present results. Section III presents the solution for time
dependent coefficients and discusses its main properties. In Sec.
IV  we discuss the possibilities to retain nonclassical features
in the output of the amplifier relating our results to earlier
calculations. In Sec. V we discuss the emergence of thermal
features in the solution. Finally Sec. VI concludes the discussion
of the work.

\section{Review on phase insensitive narrow band linear
amplifiers}\label{sec:lastandard}
The simplest standard amplifier configuration consists of an
assembly of $N$ two-level atoms, of which $N_2$ are excited and
$N_1$ are unexcited, interacting with a single-mode quantum field.
It is assumed that the field frequency is resonant with the atomic
frequency and that the population of the two-level atoms is partly
inverted, i.e. $N_2>N_1$. This is the standard model of a laser;
its linear operation regime describes an amplifier \cite{mandel}.
In the standard description of linear amplifiers it is assumed
that $N_2$ and $N_1$ are maintained approximately constant in time
by some pump and loss mechanism.

Starting from a microscopic description of the interaction between
the two-level atoms and the quantum mode, it is possible to derive
the following master equation for the field mode in the
interaction picture \cite{mandel}
\begin{eqnarray}
\frac{ d \rho}{d t}= &-& \frac{C}{2} \left[ a^{\dag} a \rho - 2 a
\rho a^{\dag} + \rho a^{\dag} a \right]
\nonumber \\
&-& \frac{A}{2} \left[  a a^{\dag} \rho - 2 a^{\dag} \rho a + \rho
a a^{\dag}
 \right], \label{MElaco}
\end{eqnarray}
with $a$ and $a^{\dag}$ annihilation and creation operator of the
quantum harmonic oscillator and
\begin{eqnarray}
A &=& \frac{2g^2}{\gamma^2} r_2, \label{Aconst}\\
C &=& \frac{2g^2}{\gamma^2} r_1, \label{Cconst}
\end{eqnarray}
where $g$ is the coupling strength of the interaction  between the
two-level atoms and the mode of the field, $r_i=N_i/\gamma$ (with
$i=1,2$) is the pumping rate into the atomic level $i$, and
$\gamma$ is a rate of the same order of the atomic linewidth.

A relevant quantity in the dynamics is the linear gain (or
damping) factor, describing the linear growth (loss) of energy in
the mode,
\begin{equation}
W=A-C.
\end{equation}
When $W>0$, the master equation describes a linear amplifier, when
$W<0$ it describes a linear absorber. The constant A gives the
noise provided by the spontaneous emission; this term is present
even if the mode energy is initially zero.

The time evolution of the amplitude of the field is described by
the equation:
\begin{equation}
\langle a \rangle_{out} \equiv \langle a(t) \rangle = G^{1/2}
e^{-i \omega_0 t} \langle a \rangle_{in},
\end{equation}
where $\langle a \rangle_{in}=\langle a(t=0)\rangle$, $\omega_0$
is the frequency of the radiation field and the gain $G$ is
defined as
\begin{equation}
G=e^{W t}. \label{gaintconst}
\end{equation}
The gain is greater than $1$ for linear amplifiers and smaller
than $1$ for linear attenuators. The solution of the Fokker-Plank
equation for the Glauber-Sudarshan $P$ representation of the
density matrix ($P$ function) can be used to calculate the
transformation of any incoming $P_{in}(\alpha)$ function by the
amplifier:
\begin{equation}
P_{\rm out} (\alpha) = \int d \alpha_0 P(\alpha,t|\alpha_0)
P_{in}(\alpha_0) \label{eq:p1}
\end{equation}
where
\begin{equation}
P(\alpha,t|\alpha_0)= \frac{1}{\pi m(t)}\exp \left[-
\frac{\left|\alpha - G^{1/2} e^{-i \omega_0 t} \alpha_0
\right|^2}{m(t)} \right]\label{eq:p2}
\end{equation}
is the amplifier transfer function \cite{stig}. The  time
dependent width is given by
\begin{equation}
m(t) = A\left[ G(t) -1 \right]/W.
\end{equation}
This quantity represents the average photon number of the
spontaneous emission field \cite{mandel}. Note that, for a linear
amplifier, the gain grows asymptotically to infinity for
$t\rightarrow \infty$, and so does the width \cite{glauber}. For
an absorber, on the other hand, the asymptotic value of the width
$m(t)$ is finite.

By using Eqs. (\ref{eq:p1}) and (\ref{eq:p2}) one can calculate
the noise of the output field, defined as the symmetrically
ordered fluctuations of the field mode, \cite{caves}:
\begin{eqnarray}
|\Delta a|^2_{out} &=& \frac{1}{2} \langle a^{\dag} a + a a^{\dag}
\rangle _{out} - \langle a\rangle_{out} \langle
a^{\dag}\rangle_{out} \nonumber \\
&=& G  |\Delta a|^2_{in}+ m(t) - \frac{1}{2}(G-1) \nonumber \\
 &\equiv& G \left( |\Delta a|^2_{in}
+ \mathcal{A}\right), \label{eqconst1}
\end{eqnarray}
where
\begin{equation}
\mathcal{A} = \frac{1}{2}\left( \frac{A+C}{A-C}\right) \left( 1-
\frac{1}{G} \right),  \label{eqconst2}
\end{equation}
is the equivalent noise factor, or amplifier added noise,
introduced by Caves \cite{caves}. This quantity describes the
fluctuations of the internal modes of the amplifying medium. Since
the input field and the internal modes of the amplifying medium
are uncorrelated, their fluctuations add in quadrature and they
are both amplified. The minimum value of the added noise for
infinite gain, e.g. for $t\rightarrow \infty$, is given by the
Caves limit:
\begin{equation}
\mathcal{A}_C = \frac{1}{2}\left( \frac{A+C}{A-C}\right) =
\frac{1}{2} + \theta \geq \frac{1}{2}, \label{eq:caves0}
\end{equation}
where the excess noise factor $\theta$ gives the initial mean
number of excitations of the internal modes of the medium, and
therefore approaches zero when the initial temperature of the
amplifying medium vanishes, $T\rightarrow 0$.

It has been demonstrated that the output field of a phase
insensitive narrow band linear amplifier may possess nonclassical
features only if the input field is nonclassical. However, as the
amplifier gain increases, any nonclassical feature of the light,
which was present in the input field, tends to be lost. In
particular, subPoissonian statistics and squeezing are lost when
the gain $G$ exceeds the value 2 \cite{mandel,stig}.

In the next section we present a theory describing the transient
regime of the amplification process. In other words, we will drop
the assumption that $N_2$ and $N_1$ are constant and we will
describe the onset of the amplification process from an initial
damping regime. Our aim is to study the transient dynamics and to
investigate how the results for the standard amplifier, described
in this section, are modified.

\section{Transient regime of linear amplification}

Previous work on linear amplifiers deals with a situation in which
the amplifying medium, e.g. an assembly of two-level atoms,
satisfies the population inversion condition required to amplify
an input field. In order to reach the inverted population
condition it is necessary to pump the atoms from their initial
thermal condition till the point in which $N_2>N_1$. During this
transient regime the pumping rates to levels 2 and 1 ($r_2(t)$ and
$r_1(t)$) change with time till they reach a stationary value for
which $N_2/N_1 \propto r_2/r_1 > 1$ (amplification regime). In
this case, the time evolution of the field mode is described by a
master equation of the same form of Eq. (\ref{MElaco}), but with
time dependent coefficients $A(t)$ and $C(t)$. Similarly, if one
switches off the external pump, the ratio of the atomic
populations $N_2/N_1$ will tend to the Boltzmann factor and the
amplification process will eventually stop, the system approaching
its thermal equilibrium. In the following we consider the first of
these two situations, i.e. the onset of amplification due to the
creation of population inversion in an initially damping medium.
In more detail, we consider the case in which
\begin{eqnarray}
A(t) &=& \frac{2g^2}{\gamma^2} r_2(t) = A
\frac{e^{\varepsilon(t-t_0)}}{e^{\varepsilon(t-t_0)}+e^{-\varepsilon(t-t_0)}} + B, \label{at} \\
C(t) &=&\frac{2g^2}{\gamma^2} r_1(t) = A \frac{e^{-
\varepsilon(t-t_0)}}{e^{\varepsilon(t-t_0)}+e^{-\varepsilon(t-t_0)}}
+ B, \label{bt}
\end{eqnarray}
where $\varepsilon$ is the rate of change of the pumping
coefficients, that is the amplification onset rate. We assume that
at $t=-\infty$ the state of the ensemble of two-level atoms
constituting the amplifying medium is thermal, that is
\begin{eqnarray}
\frac{N_2(t\rightarrow - \infty)}{N_1(t\rightarrow - \infty)}&=&
\frac{A(t\rightarrow - \infty)}{C(t\rightarrow - \infty)} =
\frac{B}{A+B}\\
&=& e^{-\hbar \omega_0 /k_BT}, \nonumber
\end{eqnarray}
which implies $B/A = (e^{\hbar \omega_0/k_BT} -1)^{-1}=n_M$, with
$n_M$ mean number of excitations of the medium. We assume that the
state of the amplifying medium practically does not change in the
time interval $- \infty < t \le 0$.

Under these conditions the asymptotic gain factor $W(t)$ takes the
form
\begin{equation}
W(t)=A(t)-C(t)= A \tanh [\varepsilon (t-t_0) ]. \label{wt}
\end{equation}
Note that, for $t < t_0 $, $W(t)<0$ and the system behaves as an
absorber, while for $t>t_0$, $W(t)>0$ and the system behaves as an
amplifier. Therefore $t_0$ indicates the time at which the
amplification process begins.
From Eq. (\ref{wt}) we infer that the constant $A$ is the
asymptotic gain factor. Finally we stress that, for $t \rightarrow
\infty$,
\begin{eqnarray}
\!\!\!\!&N_2&(t\rightarrow  \infty)/N_1(t\rightarrow  \infty)=
A(t\rightarrow  \infty)/C(t\rightarrow  \infty) \nonumber \\ &=&
(A+B)/B = N_1(t\rightarrow  - \infty)/N_2(t\rightarrow  - \infty),
\end{eqnarray}
i.e., the population of the exited (ground) state tends
asymptotically to the initial population of the ground (excited)
state.

\subsection{The master equation and its solution}
The master equation describing the transient behavior of the
amplification process, in the interaction picture, is the
following
\begin{eqnarray}
\frac{ d \rho}{d \tau}= &-& \frac{A'(\tau)}{2} \left[  a a^{\dag}
\rho
- 2 a^{\dag} \rho a + \rho a a^{\dag} \right]\nonumber \\
&-&\frac{C'(\tau)}{2} \left[ a^{\dag} a \rho - 2 a \rho a^{\dag} +
\rho a^{\dag} a \right] , \label{MEla}
\end{eqnarray}
with
\begin{eqnarray}
A'(\tau) &=&  A' \frac{e^{(\tau-\tau_0)}}{e^{(\tau-\tau_0)}+e^{-(\tau-\tau_0)}} + B', \label{eq:aprime}\\
C'(\tau) &=& A' \frac{e^{-(
\tau-\tau_0)}}{e^{(\tau-\tau_0)}+e^{-(\tau-\tau_0)}} + B'.
\label{eq:cprime}
\end{eqnarray}
In the previous equations we have introduced the relevant physical
dimensionless parameters $\tau = \varepsilon t$, $\tau_0 =
\varepsilon t_0$, $A'= A/\varepsilon$ and $B'=B/\varepsilon$. As
we will see in the following, the parameter $A'$, which is the
ratio between the asymptotic gain factor and the rate of onset of
the amplification, plays a central role in the system dynamics.
Indeed both the gain $G(\tau)$ and the added noise depend
crucially on this parameter.

Following the method developed in \cite{PRAsolanalitica} we solve
the master equation given by Eq. (\ref{MEla}) in terms of the
quantum characteristic function (QCF) \cite{barnett}, defined
through the equation
\begin{equation}
\label{sdef} \rho_S(\tau)=\frac{1}{2\pi}\int \chi_{\tau}(\xi)\:
e^{\left(\xi^* a - \xi a^{\dag}\right)} d^2 \xi.
\end{equation}
The solution reads as follows
\begin{equation}
\chi_{\tau} (\xi) = e^{- \Delta(\tau) |\xi|^2} \chi_0 \left(
G^{1/2}(\tau) e^{-i (\omega_0/\varepsilon) \tau} \xi \right),
\label{chifinale}
\end{equation}
where $\chi_0$ is the QCF of the initial state of the field,
$\omega_0$ is the field frequency and the $G(\tau)$ is the gain,
given by
\begin{equation}
G(\tau) = e^{\int_0^{\tau} W(\tau') d\tau'}. \label{gt}
\end{equation}
The quantity $\Delta(\tau)$, appearing in Eq. (\ref{chifinale}) is
defined as follows
\begin{equation}
\Delta(\tau)=\frac{1}{2} [G(\tau)] \int_0^t [G(\tau')]^{-1}
\left[C'(\tau')+A'(\tau')\right] d\tau'. \label{deltag0}
\end{equation}
It is worth underlining that the solution given by Eq.
(\ref{chifinale}), with the help of Eqs.
(\ref{gt})-(\ref{deltag0}), holds whatever the explicit time
dependence of the coefficients $A(\tau)$ and $C(\tau)$, appearing
in Eq. (\ref{MEla}), is. The case considered in the paper [Eqs.
(\ref{at})-(\ref{bt})] has been chosen to illustrate the transient
dynamics in a physically reasonable and well justified model.
Indeed Eqs. (\ref{at})-(\ref{bt}) describe a situation in which
from an initial condition in which $N_1>N_2$, the populations of
the two-level systems constituting the amplifying medium pass
smoothly to the inversion condition $N_2>N_1$ necessary for
amplification. In passing, we note that the hyperbolic tangent
time dependence is one of the most commonly adopted
phenomenological models in the description of transients of
physical systems.

Starting from Eq. (\ref{chifinale}), one can calculate the Wigner
function, the Glauber-Sudarshan $P$ function, and the Husimi $Q$
function by means of the relation \cite{barnett}
\begin{equation}
W_{\tau}(\alpha,p)= \frac{1}{\pi^2}\int^{\infty}_{-
\infty}\!\!\!\!d^2 \xi \chi_{\tau} (\xi) \exp(\alpha \xi^* -
\alpha^* \xi) e^{(p |\xi|^2/2)}, \label{eq:qpfunctions}
\end{equation}
where $p=-1,0,1$ corresponds to the $Q$, Wigner, and $P$
functions, respectively. In particular, carrying out the
calculations, it turns out that the $P$ function has the same form
of Eq. (\ref{eq:p2}), but with $G(\tau)$ given by Eq. (\ref{gt})
and $m(\tau) = \left[ G(\tau) - 1\right]/2 + \Delta (\tau)$.

Inserting Eq. (\ref{wt}) into Eq. (\ref{gt}) and carrying out the
integration yields
\begin{equation}
G(\tau)=
\left[\frac{\cosh(\tau-\tau_0)}{\cosh(\tau_0)}\right]^{A'}.
\label{gainfun}
\end{equation}
It is not difficult to prove that, for $\tau_0=0$, and in the
limit of infinitely fast onset of the amplification ($\varepsilon
\rightarrow \infty$), the gain function tends to $G(t)=e^{A
t}=e^{Wt}$ [see Eq. (\ref{gaintconst})].

\begin{figure}
\resizebox{0.50\textwidth}{!}{
\includegraphics{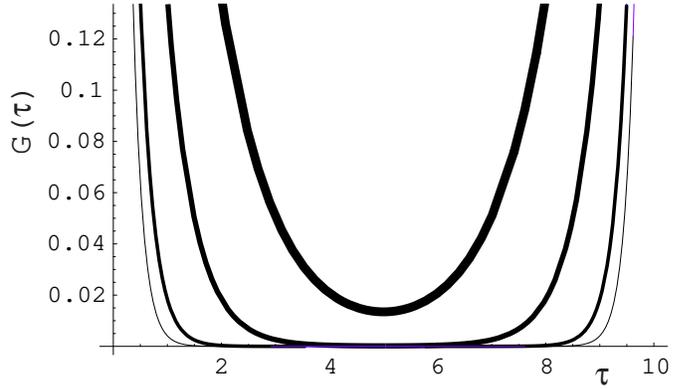}}
\caption{\label{fig:gain} Time evolution of $G(\tau)$ for
$A'=2,3,4,5.5$ (increasing values of $A'$ correspond to decreasing
thickness of the line) and $\tau_0=5$.}
\end{figure}

In Fig. \ref{fig:gain} we plot the gain $G(\tau)$ for four
increasing values of $A'$.
As clearly shown in the figure for increasing values of $A'$ the
values of the gain in proximity of the amplification time $\tau_0$
become smaller and smaller. This is because small values of $A'$
correspond to small values of the asymptotic gain factor or,
equivalently, to a very slow amplification onset rate. In general,
the gain decreases for times smaller than $\tau_0$ and, as
expected, starts to increase after the amplification sets in, even
if $G(\tau)>1$ only for times $\tau > 2 \tau_0$. Note that the
standard theory of linear amplification predicts that for a linear
amplifier it is always $G(\tau)>1$ (see Sec.
\ref{sec:lastandard}). However, if one takes into account the
transient regime characterizing the initial dynamics of every
linear amplifier it turns out that there exist an interval of time
at which, although $W>0$, there is still no gain. The reason why
the gain becomes greater than $1$ only after the time $2 \tau_0$
is that for $0<\tau<\tau_0$ the system is in a damping regime, and
hence the gain decreases. It takes exactly another interval of
time $\tau_0$ after the onset of the amplification process to undo
the initial decrease in the gain. At $\tau=2 \tau_0$ we have
$G(2\tau_0)=G(0)=1$ after which the gain increases monotonically.
This is a new feature brought to light by our theory.

Let us focus on the quantity $\Delta(\tau)$. Inserting Eq.
(\ref{gainfun}) into Eq. (\ref{deltag0}) we get
\begin{eqnarray}
\Delta(\tau)&=& G(\tau) \frac{A'+2B'}{2} \int_0^{\tau}\!\!\!
\left[\frac{\cosh(\tau'-\tau_0)
}{\cosh(\tau_0)}\right]^{-A'}\!\!\! d\tau' \nonumber \\  &\equiv&
G(\tau) \frac{A'+2B'}{2} I_{A'}(\tau-\tau_0). \label{eq:delta}
\end{eqnarray}
For $\tau_0=0$ and for each $A'$ real, with $A'\neq 1$, we have
\begin{eqnarray}
I_{A'}(\tau)&=&\frac{\sinh \tau }{(A'-1)}\frac{1}{(\cosh  \tau)
^{A'-1}} \label{eq:i2ageneral} \\ && F[1,1-A'/2,3/2-A'/2;(\cosh
\tau)^2],\nonumber
\end{eqnarray}
with $F[1,1-A'/2,3/2-A'/2;(\cosh \tau)^2]$ being the
hypergeometric function of the variable $x=(\cosh \tau)^2$. For
$A'=1$ the integral appearing in Eq. (\ref{eq:delta}) is simply
equal to
\begin{equation}
I_1(\tau)= 2 \arctan (e^{\tau}).
\end{equation}
The mathematical expression of the added noise in the special case
of integer values of $A'$ is discussed in Appendix A. For $\tau_0
\neq 0$, one gets
\begin{eqnarray}
\Delta(\tau)= G(\tau) \frac{A'+2B'}{2} \cosh (\tau_0)^{A'}
\nonumber
\\ \left[ I_{A'}(\tau-\tau_0) + I_{A'}(\tau_0) \right],
\label{deltag}
\end{eqnarray}
where $I_{A'}(\tau - \tau_0)$ and $I_{A'}(\tau_0)$ are obtained
from Eq. (\ref{eq:i2ageneral}) by substituting for the variable
$\tau$ the expressions $\tau-\tau_0$ and $\tau_0$, respectively.

\subsection{Noise of the output field}
From the QCF solution given by Eq. (\ref{chifinale}) we can easily
calculate the mean values of observables of interest, e.g. those
characterizing the output field statistics, by means of the
relation \cite{barnett}
\begin{equation}
\langle a^{\dag m} a^n \rangle = \left. \left(\frac{d}{d
\xi}\right)^m \left(- \frac{d}{d \xi^*}\right)^n
e^{|\xi|^2/2}\chi(\xi) \right|_{\xi=0}. \label{cond}
\end{equation}
We look first of all at the symmetrically ordered fluctuation, as
defined by Caves \cite{caves}
\begin{eqnarray}
|\Delta a|^2_{out} &=& G(\tau)  |\Delta a|^2_{in} + \Delta(\tau)
\nonumber \\ &=& G(\tau) \left[ |\Delta a|^2_{in} +
\mathcal{A}(\tau) \right], \label{outnoise}
\end{eqnarray}
The added noise is given by
\begin{equation}
\mathcal{A}=\Delta(\tau)/G(\tau),\label{eq:ageneral}
\end{equation}
with $\Delta(\tau)$ given by Eq. (\ref{deltag}). This quantity is
clearly different from the one given by Eq. (\ref{eqconst2}) for
the standard linear amplifier case. It is possible to show that
for infinite gain, e.g. for $\tau \rightarrow \infty$, this
quantity is always greater or equal to $\mathcal{A}_C$,
$\mathcal{A}_C$ being the Caves limit for an infinitely fast onset
of the amplification process.

In order to derive the Caves limit from Eq. (\ref{eq:ageneral}),
we note that for $\tau \rightarrow \infty$, $I_{A'}$, as given by
Eq. (\ref{eq:i2ageneral}), tends to \cite{abramowitz}
\begin{equation}
I_{A'}^{\infty} = \frac{\sqrt{\pi}}{2}
\frac{\Gamma(A'/2)}{\Gamma[(A'+1)/2]}.
\end{equation}
Therefore the asymptotic value of the added noise is
\begin{eqnarray}
&& \mathcal{A} = \frac{A'+2B'}{2} I_{A'} \rightarrow
\frac{A'+2B'}{2}\frac{\sqrt{\pi}}{2}
\frac{\Gamma(A'/2)}{\Gamma[(A'+1)/2]}\nonumber \\
&& = \left(\frac{1}{2} + \frac{B}{A}\right) \sqrt{\pi}
\frac{\Gamma(A'/2+1)}{\Gamma[(A'+1)/2]}. \label{eq:asympt}
\end{eqnarray}
The Caves limit is obtained for an infinitely fast onset of the
amplification process, that is for $\varepsilon \rightarrow
\infty$, i.e. $A' \rightarrow 0$. Substituting $A' = 0$ into Eq.
(\ref{eq:asympt}), and remembering that $\Gamma(1)=1$ and
$\Gamma(1/2)=\sqrt{\pi}$, one gets
\begin{equation}
\mathcal{A}_C = \frac{1}{2} + \frac{B}{A} \ge \frac{1}{2},
\end{equation}
that is the Caves limit [see Eq. (\ref{eq:caves0})]. From Eq.
(\ref{eq:asympt}) one can see that, for fixed values of the
initial mean number of excitations of the medium $n_M=B/A$,
$\mathcal{A}_C$ is actually the smallest asymptotic value of the
added noise.

\begin{figure}
\resizebox{0.50\textwidth}{!}{
\includegraphics{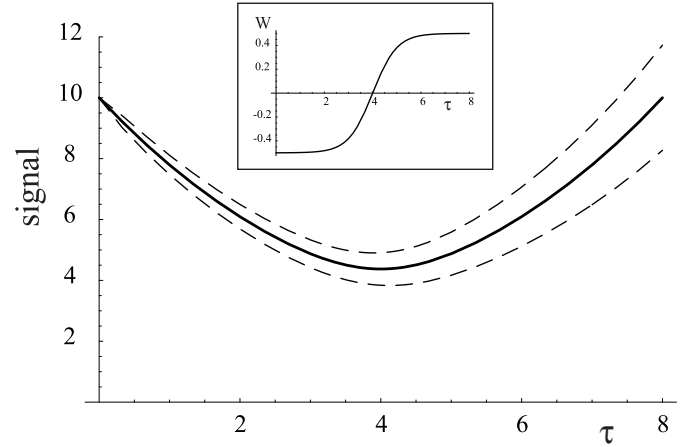}}
\caption{\label{fig:signal} Damping and amplification of an
initial input field having $\langle a(t=0) \rangle$=10, for
$\tau_0=4$, $A' = 0.5$ and $B' =0.5 \cdot 10^{-2}$. The dashed
lines indicate signal width $\Delta(\tau)^{1/2}$. The insert shows
the gain factor $W(\tau)=A(\tau)-C(\tau)$ in the same time
interval and for the same values of the parameters.}
\end{figure}

A careful analysis of the noise at the output field, as given by
Eq. (\ref{outnoise}), shows that this quantity, as one would
expect, increases monotonically with time whatever the initial
state is. In Fig. \ref{fig:signal}  we show the dynamics of an
input field which is damped for $\tau<\tau_0=4$ and then amplified
for $\tau>\tau_0=4$. The figure shows that the width of the signal
always increases.
In the following section we will study in more detail the
transient dynamics for different types of input fields. Since we
are dealing with phase insensitive amplifiers/absorbers, it turns
out that it is not possible to generate nonclassical states from
classical input fields. However, it is possible to analyze how the
conditions to retain initial nonclassical features are modified
due to the transient dynamics. In addition we will look at the
field statistics by explicitly calculating the time evolution of
the Wigner function.

\section{Nonclassical properties of the output field}
\subsection{Squeezing and subPoissonian statistics}
\label{sec:squeez} Let us begin studying the conditions for which
the output field can retain squeezing properties when the input
state is a squeezed state of the electromagnetic field. We define
the dimensionless quadratures of the field as follows
\begin{eqnarray}
u &=& \frac{1}{\sqrt{2}}\left(a+a^{\dag} \right),  \\
v &=& \frac{-i}{\sqrt{2}}\left(a-a^{\dag} \right).
\end{eqnarray}
The squeezed states satisfy the minimum uncertainty relation
$\Delta u \Delta v = 1/2$, but are characterized by an unequal
distribution of the quantum fluctuations
\begin{eqnarray}
\Delta u = \frac{s}{\sqrt{2}}, \hspace{1cm}  \Delta v =
\frac{1}{\sqrt{2} s},
\end{eqnarray}
with $s \neq 1$. Introducing the rotating coordinates
\begin{eqnarray}
\tilde{u}&=& u \cos(\omega_0 t) - v \sin (\omega_0 t),\\
\tilde{v}&=& v \cos(\omega_0 t) + u \sin (\omega_0 t),
\end{eqnarray}
and using Eq. (\ref{cond}) we get
\begin{eqnarray}
(\Delta \tilde{u})^2_{out}= G(\tau) \left[(\Delta
\tilde{u})^2_{in} + \mathcal{A}\right], \label{eq:deltau} \\
(\Delta \tilde{v})^2_{out}= G(\tau) \left[(\Delta
\tilde{v})^2_{in} + \mathcal{A}\right], \label{eq:deltav}
\end{eqnarray}
with $\mathcal{A}$ given by Eqs. (\ref{eq:ageneral}) and
(\ref{eq:delta}). For an input squeezed state having $s<1$, the
output can remain squeezed if and only if
\begin{equation}
G(\tau) \left[(\Delta \tilde{u})^2_{in} +
\mathcal{A}\right]<\frac{1}{2}.
\end{equation}
It is possible to show that the maximum allowed value of the gain
$G(\tau)$, in order to retain squeezing at the output, decreases
with $A'$ and $B'$. In other words, for increasing values of $A'$
and $B'$, one can retain squeezing only for smaller and smaller
values of the gain (less efficient amplification). Having in mind
Eq. (\ref{deltag}), one finds that, for $\tau_0=0$, the output
field is still squeezed if the gain satisfies the following
inequality
\begin{equation}
G(\tau) < \frac{1}{s^2+(A'+2B')I_{A'}(\tau)}.
\label{eq:inesqueezing}
\end{equation}
It is worth recalling the standard result for phase insensitive
linear amplifiers, which states that the upper limit for the gain
compatible with squeezing at of the output field is $G=2$, the
magic number for photon cloning \cite{stig}. The analysis of the
behavior of the gain in our case is more complicated, since the
r.h.s. of the inequality (\ref{eq:inesqueezing}) depends on time.
A numerical study shows that, although for certain time intervals,
the r.h.s of the inequality may be greater than 2, in these time
intervals $G(\tau)$ is always smaller than $2$. Hence, also in the
case studied in this paper $G=2$ constitutes an upper limit for
retaining nonclassical features in the output field.

\begin{figure}
\resizebox{0.50\textwidth}{!}{\includegraphics{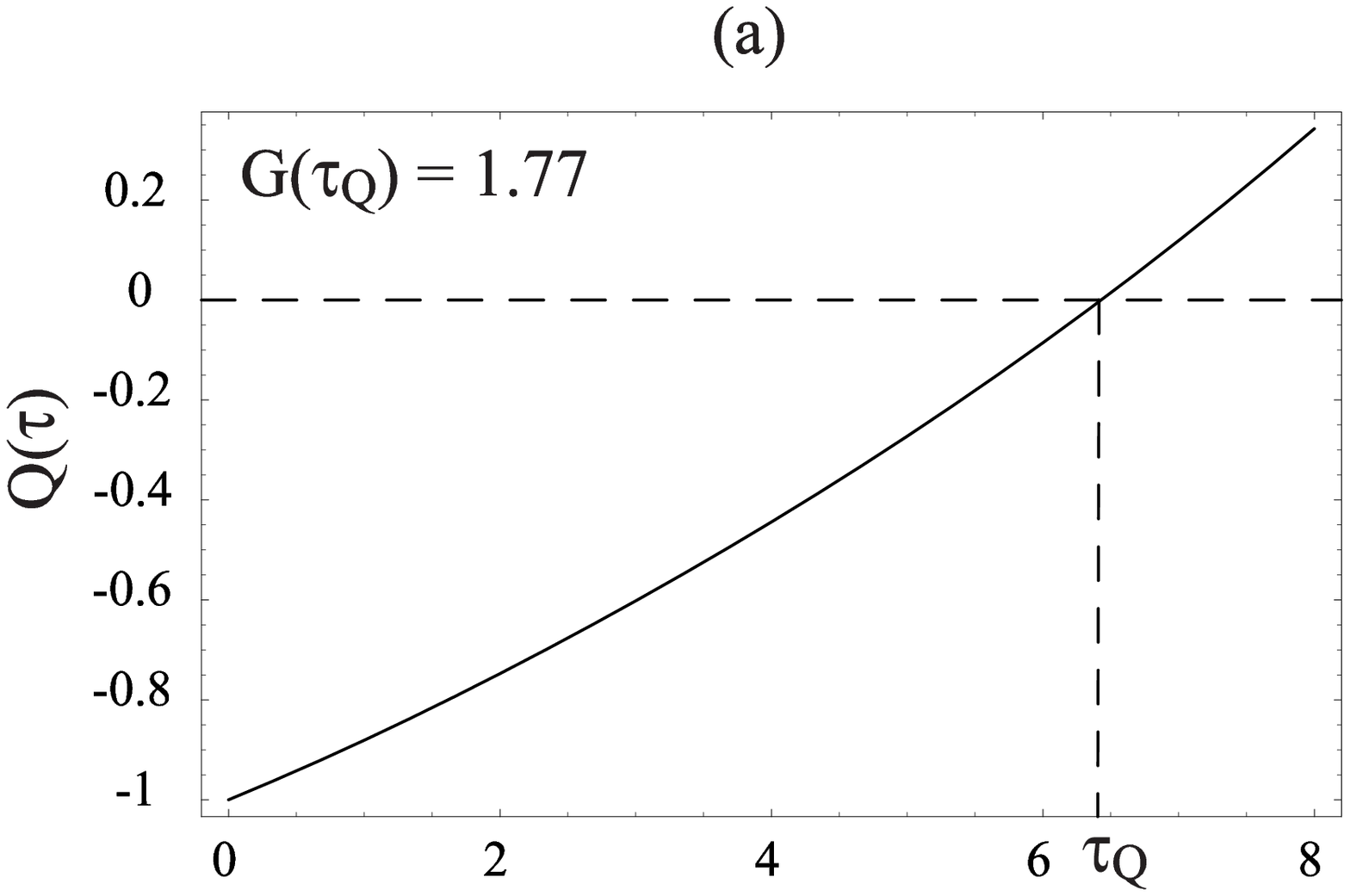}}
\resizebox{0.50\textwidth}{!}{\includegraphics{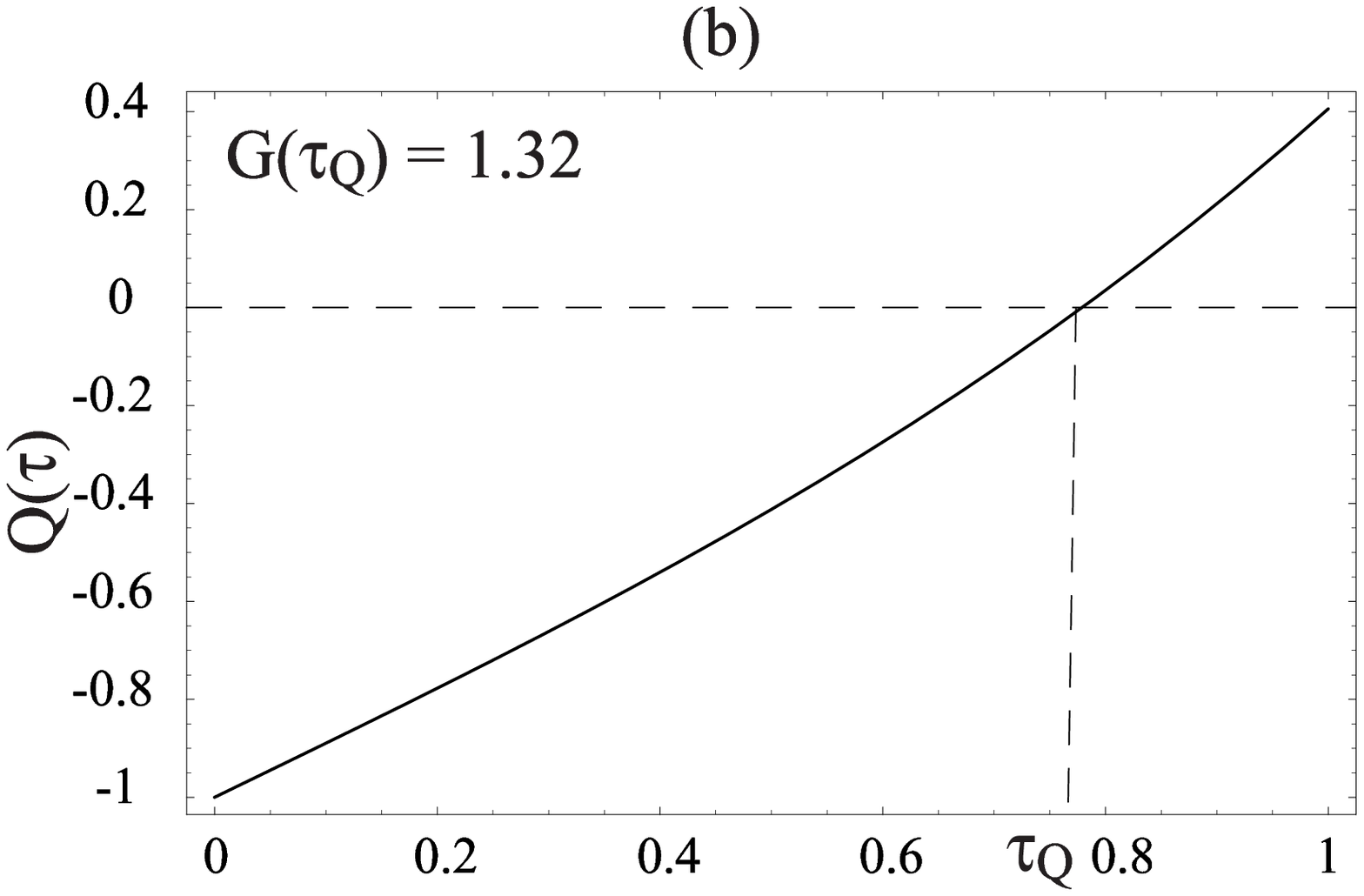}} \caption{
Mandel parameters of the output fields for an initial Fock state
$\vert n_0=5\rangle$ in the cases $A'=0.05$ (a), and $A' = 1$ (b).
We have set $\tau_0=0$ and $B/A=10^{-2}$ in both (a) and (b). We
indicate with $\tau_Q$ the instant of time at which $Q=0$. In the
upper-left corner we indicate the corresponding value of the gain
at $\tau=\tau_Q$: $G(\tau_Q)$. }\label{fig:Q}
\end{figure}

Let us now look at the dynamics of an input Fock state. We recall
that one of the nonclassical features of such states is that their
statistics is subPoissonian. Similarly to what we have done for
the squeezed states we analyze the requirements to retain
subPoissonian statistics at the output field. To this aim we
introduce the Mandel parameter $Q$ \cite{mandel}
\begin{equation}
Q=\frac{\langle n^2  \rangle  - \langle n \rangle ^2}{\langle n
\rangle}-1.
\end{equation}
This quantity gives an indication of the statistics of a quantized
field. For a Fock state  $Q$ takes its lowest value $Q=-1$ while
for a coherent state $Q$ is equal to $0$ . Therefore, values of
$Q<0$ indicate subPoissonian statistics, while $Q = 0$ indicates
Poissonian statistics and $Q>0$ superPoissonian statistics.  Using
Eq. (\ref{cond}) we derive the time evolution of the Mandel
parameter as follows
\begin{eqnarray}
Q_{out}\!=\!\frac{\langle n \rangle^2_{out} \!\!+ [G(\tau)]^2 \!
\langle n \rangle _{in}  \!\left[  Q_{in} -\! \langle n \rangle
_{in}\right]}{\langle n \rangle _{out}}, \label{eq:Q}
\end{eqnarray}
where $\langle n \rangle _{out}$ and $\langle n \rangle _{in}$ are
the mean number of photons of the output and input fields,
respectively, and
\begin{eqnarray}
\langle n \rangle_{out} &=& G(\tau) \langle n \rangle _{in}+
\frac{1}{2} \left[  G(\tau)  - 1\right] + \Delta(\tau)
\nonumber \\
&=& G(\tau) \langle n \rangle _{in}+ m(\tau). \label{eq:nout}
\end{eqnarray}
For a Fock input state $Q_{in}=-1$ and $\langle n \rangle
_{in}=n_0$, hence the condition for having subPoissonian
statistics at the output is
\begin{equation}
\langle n \rangle_{out}(\langle n \rangle_{out}+1) < G^2(\tau)
n_0(n_0+1).
\end{equation}
A numerical analysis shows that, as for the squeezing, also for
the subPoissonian statistics, we obtain the limit $2$ for the gain
typical of  the standard theory of linear amplifiers.

As an example, in Fig. \ref{fig:Q} we compare the Mandel
parameters of the output fields for an initial Fock state $\vert
n_0=5\rangle$ in the cases $A'=0.05$ (fast onset of the
amplification and/or small value of the asymptotic gain) and
$A'=1$. From the figure one sees that, for $A'=0.05$, one can
retain subPoissonian statistics of the output field for higher
values of the gain compared to the $A' = 1$ case.

\subsection{Wigner function}
Let us now have a look at the complete statistic of the output
field, by means of the Wigner function. Inserting Eq.
(\ref{chifinale}) into Eq. (\ref{eq:qpfunctions}), and putting
$p=0$ we get
\begin{eqnarray}
W_{\tau}(\alpha)=\frac{1}{\pi^2}\int_{-\infty}^{\infty}\!\!\!\!
d^2 \xi \,
e^{- \Delta(\tau)|\xi|^2} \, e^{\alpha \xi^* - \alpha^* \xi} \nonumber\\
\chi_0 (G^{1/2}(\tau) e^{-i \omega_0 \tau / \varepsilon} \xi).
\label{w1}
\end{eqnarray}
Inserting the inverse Fourier transform of Eq.
(\ref{eq:qpfunctions}) into Eq.(\ref{w1}) gives
\begin{eqnarray}
&& W_{\tau}(\alpha)= \frac{1}{\pi^2} \int_{-\infty}^{\infty} \,
d^2 \alpha_0 W_0(\alpha_0) \nonumber \\&&  \int_{-\infty}^{\infty}
\, d^2\xi \, e^{- \Delta(\tau)|\xi|^2} \, e^{b(\tau, \alpha ,
\alpha_0)
\xi^* - b^*(\tau, \alpha , \alpha_0) \xi} \nonumber \\
&&= \frac{1}{\pi} \int_{-\infty}^{\infty} \!\! d^2 \alpha_0
W_0(\alpha_0) \frac{ \exp  \left[  - \frac{|b(\tau, \alpha ,
\alpha_0 )|^2}{
\Delta(\tau)} \right] }{ \Delta(\tau)} \nonumber \\
&&\equiv \frac{1}{\pi} \int_{-\infty}^{\infty} \!\! d^2 \alpha_0
W_{\tau}(\alpha|\alpha_0) \!\! W_0(\alpha_0), \label{wfinale}
\end{eqnarray}
with
\begin{equation}
b(\tau, \alpha , \alpha_0) = \alpha - \alpha_0 G^{1/2}(\tau) e^{i
\omega_0 \tau / \varepsilon}.
\end{equation}
In the derivation of Eq. (\ref{wfinale}) we have used the property
that the Fourier transform of a Gaussian is a Gaussian. The
quantity $W_{\tau}(\alpha|\alpha_0)$ is the propagator which, for
$\tau \rightarrow 0$, tends to the delta function $\delta (\alpha
- \alpha_0)$.

If the state of the input field is a coherent state
$|\alpha_0\rangle$, than the Wigner function of the output state
reads as follows
\begin{equation}
W_{\tau}(\alpha)=\frac{1}{\pi} \frac{\exp \left[ - \frac{\left|
\alpha_0 G^{1/2}(\tau) e^{i \omega_0 \tau/\varepsilon} - \alpha
\right|^2}{\Delta(\tau)+1/2}\right]}{\Delta(\tau)+1/2}.
\label{eq:wigalpha}
\end{equation}
The Wigner function of the output state is therefore a Gaussian.
Having in mind the time evolution of $G(\tau)$ [see
Eq.(\ref{gainfun}) and Fig.\ref{fig:gain}], one realizes that, in
a frame rotating with the frequency $\omega_0$, the Wigner
function of an input coherent state $\vert \alpha_0 \rangle$, with
$\alpha_0 \ne 0$, moves towards the center of the phase space for
$\tau<\tau_0$ and then moves away for $\tau>\tau_0$, while its
width continuously increases.
\begin{figure}
\resizebox{0.40\textwidth}{!}{
\includegraphics{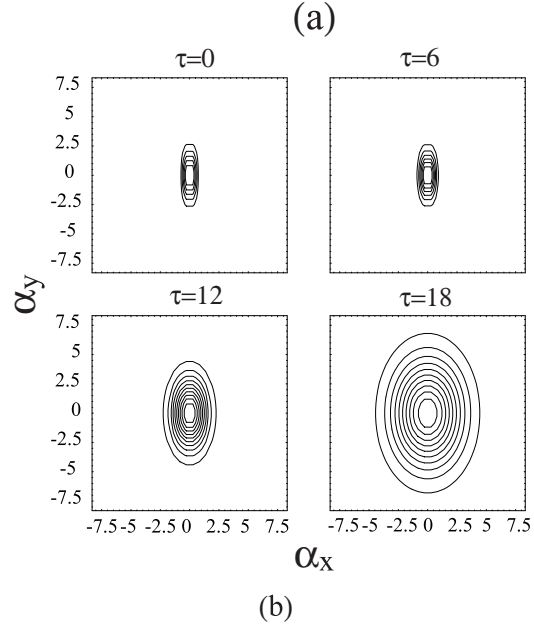}}
\resizebox{0.40\textwidth}{!}{\includegraphics{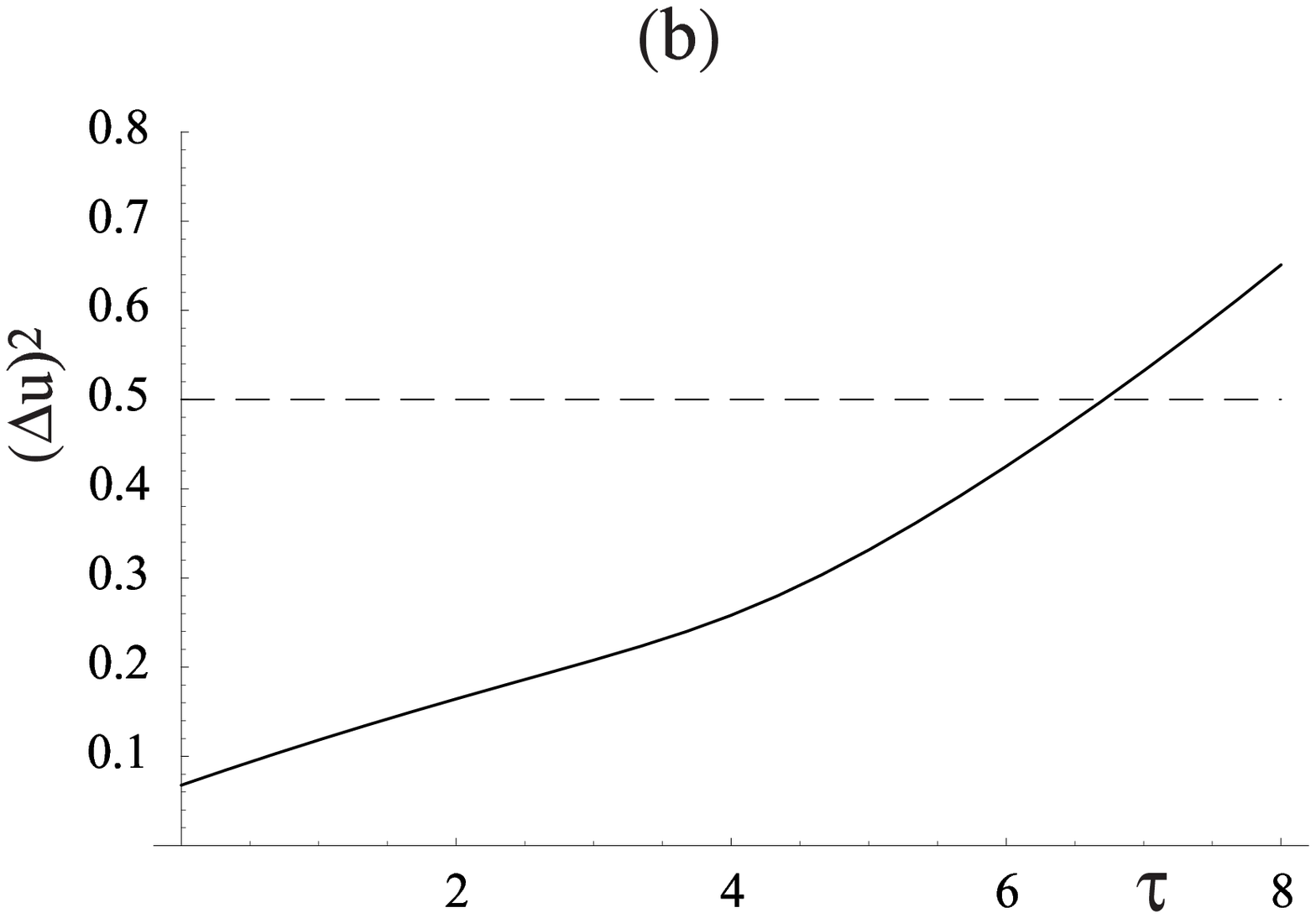}}
 \caption{ Contour plots of the Wigner function at different times $\tau$ for an input squeezed state with $r=1$.
 We have put $\tau_0=4$, $A'=0.1$, $n_B=B/A=0.1$ (a). Variance of the quadrature $\tilde{u}$ as a function of time,
 for the same values of the parameters (b).}\label{fig:wifsqueezed}
\end{figure}

We now consider the case of an initially squeezed state. The
initial QCF for squeezed coherent state is
\begin{equation}
\chi_0(\xi)= \exp\left[ -\frac{1}{2} |\xi
C_r-\xi^*e^{-i\phi}S_r|^2 +i(\xi^*\alpha_0^*+\xi\alpha_0)\right].
\end{equation}
Here $C_r=\cosh(r)$ and $S_r=\sinh (r)$ , $\alpha_0$ is the
displacement of the input field and $z=re^{-i\phi}$ is the
squeezing argument.

For an input squeezed vacuum state ($\alpha_0=0$), with squeezing
angle $\phi=0$, the Wigner function at time $\tau$ takes the form
\begin{eqnarray}
&& W_{\tau}(\alpha)= \frac{1}{\pi^2} \int_{-\infty}^{\infty} \,
d^2 \xi \,
e^{-\Delta(\tau) |\xi|^2} e^{\left( \alpha\xi^*-\alpha^*\xi\right)} \nonumber \\
&& \exp\left[ -\frac{1}{2} G(\tau)|e^{-i\omega_0 \tau/ \varepsilon
} \xi C_r   -e^{i\omega_0 \tau/ \varepsilon}\xi^*S_r|^2\right] .
\end{eqnarray}

This Fourier transformation can be calculated with the method used
in \cite{matsuo}, the result being
\begin{eqnarray}
W_{\tau}(\alpha) &=& M \exp\left[ \frac{-2
\alpha_x^2}{2\Delta(\tau)+
G(\tau)(C_{2r}+S_{2r})^{-1}}\right]\nonumber
\\ &+& \exp\left[ \frac{-2 \alpha_y^2}{2\Delta(\tau)
G(\tau)(C_{2r}-S_{2r})^{-1}}\right]
\end{eqnarray}
Here, $\alpha_x$ and $\alpha_y$ are the real and imaginary parts
of $\alpha$, and $M$ is a time dependent normalization constant.
We note that this result is consistent with the Eqs.
(\ref{eq:deltau})-(\ref{eq:deltav}) used in Sec. \ref{sec:squeez}
to study the time evolution of the quadratures of the field for an
initial input squeezed state. Contour plots showing the time
evolution of the Wigner function are shown in
Fig.~\ref{fig:wifsqueezed} (a), while in Fig.
~\ref{fig:wifsqueezed} (b) we show the time evolution of the
squeezing of the quadrature amplitude $\tilde{u}$.

\section{Departure from thermal equilibrium}
The analytic approach we have described in the previous section to
analyze the onset of the amplification process, can be used to
study how a system departs from an initial thermal equilibrium
situation. In more detail, we consider the case in which the
medium, modelled as an ensemble of two-level atoms, is initially
in thermal equilibrium with the mode of the quantized field. The
ratio $N_2/N_1$ between the populations of the excited and ground
states, respectively, is therefore given by the Boltzmann factor
$N_2/N_1 = e^{-\hbar \omega_0/k_{B}T}$. The state of the field is
a thermal state at $T$ temperature.

At $\tau=0$ one switches on pumping lasers which change the
population of the two-level atoms until the condition of
population inversion, necessary for the onset of the amplification
process, is reached. The pumping lasers alter the initial
condition of equilibrium between the medium and the system (the
field mode). In order to study how the system departs from the
condition of thermal equilibrium with the two-level atoms medium,
we use the solution of the Master Equation (\ref{MEla}) to
calculate the time evolution of the Wigner function of the field.
For an initial thermal state, the QCF at time $\tau$ has the form
\begin{equation}
\chi_{\tau}(\xi)=e^{-\Delta(\tau)|\xi|^2} \exp \left[ -\left(
\langle n\rangle _{out} +\frac{1}{2} \right) G(\tau) |\xi|^2
\right].
\end{equation}
Inserting this equation into Eq. (\ref{eq:qpfunctions}) one gets
the following expression for the Wigner function at time $\tau$
\begin{equation}
W_{\tau}(\alpha)=\frac{1}{\pi} \frac{1}{\langle n
\rangle_{out}+1/2} \exp{\left[-\frac{|\alpha|^2}{\langle n
\rangle_{out}+1/2}\right]}. \label{eq:wthermal}
\end{equation}

\begin{figure}
\resizebox{0.50\textwidth}{!}{
\includegraphics{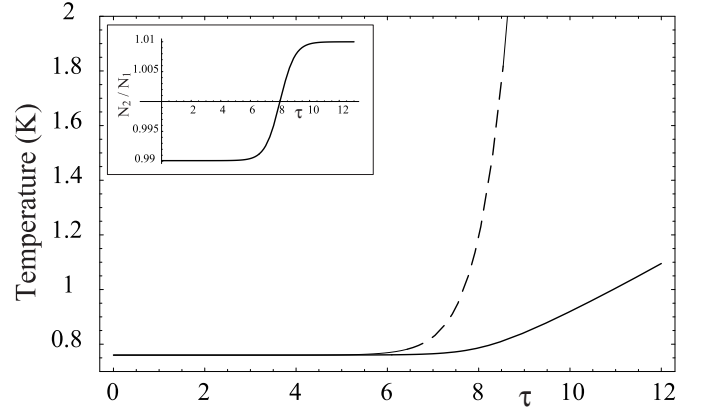}}
 \caption{ Time evolution of the temperature of
the system for $A' =1$ (dotted line) and $A'=0.05$ (solid line).
For both graphics we have set $\omega_0=10^{14} Hz$ and
$n_B=B/A=10^3$. The box in the top left corner is the ratio
between the populations of the two-level atoms. This quantity does
not depend on $A'$ and $B'$ separately, but on the ratio $n_B=B/A$
only [see Eqs. (\ref{at}) and (\ref{bt})].} \label{fig:temp}
\end{figure}

In the last two equations, $\langle n \rangle_{out}$ is the number
of photons of the output field, as given by Eq. (\ref{eq:nout}).
The Wigner function of Eq. (\ref{eq:wthermal}) is the Wigner
function of a thermal state at a temperature $T (\tau)$ which
varies with time. The medium is not in thermal equilibrium
anymore, since the pumping lasers change the two-level atoms
population until the population inversion condition is reached.
However, Eq.(\ref{eq:wthermal}) shows that  the medium plus the
pumping lasers behave, as far as the system (mode field) is
concerned, as a thermal reservoir at varying temperature
$T(\tau)$, as known from the theory of laser cooling \cite{stig2}.
Having in mind that $2 \langle n \rangle_{out} + 1 = \coth \left[
\hbar \omega_0 / k_B T(\tau) \right]$ \cite{mandel}, and using the
relation ${\rm arcoth} (x)=[\ln (x+1) - \ln (x-1)]/2$ (for
$x^2>1$), we can express the time evolution of the temperature as
follows:
\begin{equation}
T(\tau)= \frac{\hbar \omega_0}{k_B} \left\{\ln \left[ \langle n
\rangle_{out}+1 \right] -  \ln \left[ \langle
n\rangle_{out}\right] \right\}^{-1}.
\end{equation}
Figure \ref{fig:temp} compares the behavior of $T(\tau)$ for the
two cases $A'=1$ and $A'=0.05$. The figure shows clearly that in
both cases the temperature of the system is constant during the
damping regime and it starts to increase when approaching the
population inversion at $\tau=\tau_0$, i.e. when the change in the
population ratio $N_2/N_1$ becomes considerable. The increase in
the temperature is much higher for higher values of $A'$, since in
this case the asymptotic gain is also higher.

\begin{figure}
\resizebox{0.50\textwidth}{!}{
\includegraphics{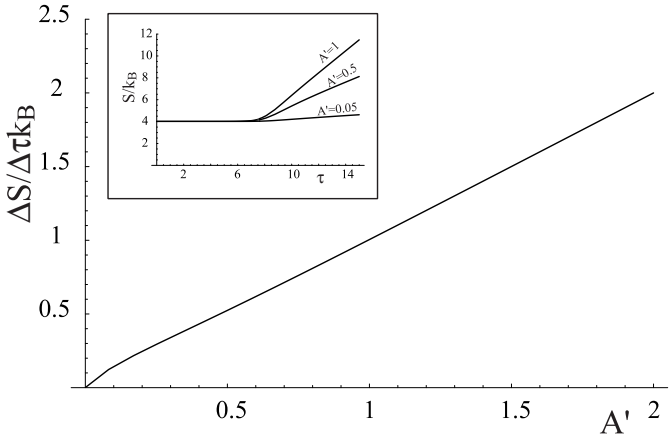}}
 \caption{ Dependence of the entropy increase
rate on $A'$. The entropy increase rate is defined as $\Delta
S/\Delta \tau k_B=\left[S(\tau=14)-S(\tau=10) \right]/4 k_B$.  In
the box in the top left corner the time evolution of the entropy
for the three exemplary values $A'=1$, $A'=0.5$, and $A'=0.05$ is
shown. In all the plots we have set $\tau_0=8$, and
$n_B=B/A=10$.}\label{fig:entropy}
\end{figure}

In order to characterize further the departure from the initial
condition of the system we look at the von-Neumann entropy of the
field. We use the result obtained by Agarwal \cite{agarwal71} to
calculate the dynamical entropy for a state of the form given by
Eq.(\ref{eq:wthermal})
\begin{eqnarray}
S(\tau)&=&k_B \left\{ \left[ \langle n\rangle_{out} + 1\right] \ln
\left[ \langle n \rangle_{out}+1 \right] \right. \nonumber
\\ &-& \left.\langle n\rangle_{out} \ln \left[ \langle
n\rangle_{out}\right] \right\}.
\end{eqnarray}
From direct inspection in the previous equation one sees that,
similarly to the dynamics of the temperature, the entropy remains
approximately constant for $\tau < \tau_0$ and begins to increase
when $\tau \simeq \tau_0$. In Fig. \ref{fig:entropy} we show how
the entropy increase rate (in units of $k_B$) changes with $A'$;
in the box in the top left corner the dynamics of the entropy for
three example values of $A'$ is shown. For increasing values of
$A'$, the linear increase in the entropy due to the amplification
process is faster and faster. This result is in accordance with
the behavior of the temperature of the system. In fact, smaller
values of $A'$ correspond to smaller asymptotic gain and therefore
less efficient amplification processes.

\section{Conclusions}
In this paper we have discussed the dynamics of a quantum linear
amplifier during the onset of the amplification process. For an
amplifying medium consisting of an assembly of two-level atoms,
our theory describes the dynamics of the output field when the
medium passes from a condition in which the population of the
atoms is thermal, to a condition of population inversion
characterizing the amplifying regime.

We have solved exactly the master equation describing the
transient dynamics of the linear amplifier in terms of the quantum
characteristic function. The solution is used to investigate
conditions under which an input nonclassical field may retain
nonclassical features at the output of the linear amplifier. We
derive the analytic expressions for the output noise, as well as
for the squeezing, the Mandel parameter and the Wigner function of
the output field, and we use them to characterize completely the
transient dynamics of the output field.

Our results are compared with earlier theories of phase
insensitive linear amplifiers which rely on the assumption that
the population inversion is instantaneously reached, i.e.
neglecting the transient regime. We show that also for a slow
onset of amplification, the gain $G(\tau)$ has to be smaller than
$2$ (the cloning magic number) in order for the output field to
retain initial nonclassical properties.

We conclude the paper analyzing the situation in which the initial
mode of the field and the two-level atoms medium are in thermal
equilibrium at $T$ temperature. An external laser pumps up the
atoms of the medium till the condition of population inversion is
reached. This is an example of dynamic departure from a thermal
equilibrium condition that can be studied analytically. We analyze
the time evolution of the temperature and of the von-Neumann
entropy on the characteristic parameters of the linear amplifier.
We find that, as known from the theory of laser cooling, the
medium plus the pumping lasers behave, as far as the system is
concerned, as a thermal reservoir at varying temperature. We find
that the entropy increase rate depends crucially on the asymptotic
gain.

\section{Acknowledgements}
This work has been supported by the European Union's Transfer of
Knowledge project CAMEL (Grant No. MTKD-CT-2004-014427). J.P. and
S.M. thank Nikolay Vitanov for the hospitality at the University
of Sofia and Stig Stenholm for the hospitality during the visit to
KTH in Stockholm. J.P. acknowledges financial support from the
Academy of Finland (project 204777), and from the Magnus Ehrnrooth
Foundation. S.M. acknowledges financial support from the Angelo
Della Riccia Italian National Foundation.

\section*{Appendix A}
For integer values of $A'$ the integral $I_{A'}$ defined in Eq.
(\ref{eq:delta}) is given by \cite{tavole}
\begin{eqnarray}
&&I_{2m}(\tau ) =\frac{\sinh \tau }{2m-1}\frac{1}{(\cosh \tau)
^{2m-1}} \nonumber \\  && \left[ 1
+\sum_{k=1}^{m-1}\frac{\Gamma(m) \Gamma(m-k-1/2)}{\Gamma(m-k)\Gamma(m-1/2)}(\cosh \tau)^{2k} \right], \label{eq:I2m} \\
&&I_{2m+1}(\tau ) =\frac{\sinh \tau }{2m}\frac{1}{(\cosh
\tau)^{2m} } \nonumber\\
&& \left[ 1+\sum_{k=1}^{m-1}\frac{\Gamma(m-k)\Gamma(m+1/2)}{\Gamma(m) \Gamma(m-k+1/2)}%
(\cosh \tau)^{2k} \right] \nonumber
\\
&&+ \frac{(2m-1)!!}{(2m)!!}\arctan (\sinh \tau).\label{eq:I2m1}
\end{eqnarray}
In this appendix we show that the two equations written above are
special cases of Eq. (\ref{eq:i2ageneral}).

For $A'= 2m \ge 1$ an even integer, the hypergeometric function
reduces to a polynomial of order $1-A'/2$
\begin{equation}
F[-m,b,c;z]=\sum_{k=0}^m \frac{(-m)_k(b)_k}{(c)_k} \frac{z^k}{k!},
\label{eq:property}
\end{equation}
where
\begin{equation}
(z)_k = \frac{\Gamma(z+k)}{\Gamma(z)}; \hspace{1cm} (z)_0=1.
\end{equation}
Using Eq. (\ref{eq:property}) and the following properties
\begin{eqnarray}
\Gamma(z+1)=z \Gamma(z) \nonumber \\
\Gamma(-z)=\frac{\pi \csc (\pi z)}{-z \Gamma(z)}, \nonumber
\end{eqnarray}
equation (\ref{eq:i2ageneral}) reduces to Eq.(\ref{eq:I2m}).

For $A'= (2m+1)$, we obtain Eq. (\ref{eq:I2m1}) from Eq.
(\ref{eq:i2ageneral}) by using the properties
\begin{eqnarray}
&-& i \sinh (\tau) F[1,1-A'/2,3/2-A'/2;(\cosh
\tau)^2] \nonumber\\&=&F[1/2-A'/2,1/2,3/2-A'/2;(\cosh \tau)^2]\nonumber \\
&=& F[-m,1/2,1-m;(\cosh \tau)^2], \nonumber
\end{eqnarray}
and
\begin{eqnarray}
&& F[-m,1/2,1-m;(\cosh \tau)^2] \\ && = \frac{1}{2^m}
\Gamma(1-m)(- \cosh \tau)^m P^m_m (i \sinh \tau),\nonumber
\end{eqnarray}
where
\begin{equation}
P^m_m(z)=(z^2-1)^{m/2} \frac{d^m}{dz^m} P_m(z),
\end{equation}
with $P_m(z)$ Legendre polinomials.

\end{document}